\documentstyle[amsmath,amsfonts]{article}
\bibliography{plain}
\title{Concurrence Vectors in Arbitrary Multipartite Quantum Systems}
\vspace{20mm}
\author{
  S. J. Akhtarshenas
\thanks{E-mail:akhtarshenas@phys.ui.ac.ir}
\\
\\
{\small Department of Physics, University of Isfahan, Isfahan,
Iran }
\\
 {\small Institute for Studies in Theoretical
Physics and Mathematics,
 Tehran 19395-1795, Iran.}}
\begin{document}
\maketitle \vspace{15mm}
%\newpage
\begin{abstract}
For a given pure state of  multipartite system, the concurrence
vector is defined by employing the defining representation of
generators of the corresponding  rotation groups. The norm of a
concurrence vector is considered as a measure of entanglement. For
multipartite pure state, the concurrence vector is regarded as
the direct sum of concurrence subvectors in the sense that each
subvector is associated with a pair of particles. It is proposed
to use the norm of each subvector as the contribution of the
corresponding pair in entanglement of the system.

{\bf Keywords: Quantum entanglement, Concurrence vector,
Orthogonal group}

{\bf PACS Index: 03.65.Ud }
\end{abstract}
%\pagebreak

%\vspace{7cm}

\section{Introduction}

Quantum entanglement, as the most intriguing features of quantum
mechanics, has been investigated for decades in relation with
quantum nonseparability and the violation of Bell's inequality
\cite{EPR,shcro,bell}.  In the last decade it has been regarded as
a valuable resource for quantum communications and information
processing \cite{ben1,ben2,ben3}, so, as with other resources such
as free energy and information, quantification of entanglement is
necessary to understand and develop the theory.

From the various measures proposed to quantify entanglement, the
entanglement of formation has been widely accepted which in fact
intends to quantify the resources needed to create a given
entangled state \cite{ben3}. In the case of pure state, if the
density matrix obtained from the partial trace over other
subsystems is not pure the state is entangled. Consequently, for
the pure state  $\left|\psi\right>$ of a bipartite system,
entropy of the density matrix associated with either of the two
subsystems is a good measure of entanglement
\begin{equation}\label{Epsi}
E(\psi)=-\rm{Tr}(\rho_A\log_2\rho_A)=-\rm{Tr}(\rho_B\log_2\rho_B),
\end{equation}
where $\rho_A=\rm{Tr}_B(|\psi\rangle\langle\psi|)$ and $\rho_B$ is
defined similarly. Due to classical correlations existing in the
mixed state each subsystem can have non-zero entropy even if
there is no entanglement, therefore the von Neumann entropy of a
subsystem is no longer a good measure of entanglement. For a
mixed state, the entanglement of formation (EoF) is defined as the
minimum of an average entropy of the state over all pure state
decompositions of the state \cite{ben3},
\begin{equation}\label{EoF1}
E_f(\rho)=\min\sum_i p_i E(\psi_i).
\end{equation}
Although the entanglement of formation has most widely been
accepted as an entanglement measure, there is no known explicit
formula for the EoF of a general state of bipartite systems
except for $2\otimes 2$ quantum systems \cite{woot} and special
types of mixed states with definite symmetry such as isotropic
states \cite{terhal} and Werner states \cite{vollb}. Remarkably,
Wootters  has shown that EoF of a two qubit mixed state $\rho$ is
related to a quantity called concurrence as \cite{woot}
\begin{equation}\label{EoF2}
E_f(\rho)=H\left(\frac{1}{2}+\frac{1}{2}\sqrt{1-C^2}\right),
\end{equation}
where $H(x)=-x\ln{x}-(1-x)\ln{(1-x)}$ is the binary entropy and
concurrence $C(\rho)$ is defined by
\begin{equation}\label{concurrence}
C(\rho)=\max\{0,\lambda_1-\lambda_2-\lambda_3-\lambda_4\},
\end{equation}
where the $\lambda_i$ are the non-negative eigenvalues, in
decreasing order, of the Hermitian matrix
$R\equiv\sqrt{\sqrt{\rho}{\tilde \rho}\sqrt{\rho}}$ and
\begin{equation}\label{rhotilde}
{\tilde \rho}
=(\sigma_y\otimes\sigma_y)\rho^{\ast}(\sigma_y\otimes\sigma_y),
\end{equation}
where $\rho^{\ast}$ is the complex conjugate of $\rho$ when it is
expressed in a standard basis such as $\{\left|11\right>,
\left|12\right>,\left|21\right>, \left|22\right>\}$ and $\sigma_y$
represents the Pauli matrix in a local basis $\{\left|1\right>,
\left|2\right>\}$. Furthermore, the EoF is monotonically
increasing function of the concurrence $C(\rho)$, so one can use
concurrence directly as a measure of entanglement. For pure state
$\left|\psi\right>=a_{11}\left| 11\right>+
a_{12}\left|12\right>+a_{21}\left|21\right>+a_{22}\left|22\right>$,
the concurrence takes the form
\begin{equation}\label{cpsi}
C(\psi)=|\langle\psi |\tilde{\psi}\rangle|
=2\left|a_{11}a_{22}-a_{12}a_{21}\right|.
\end{equation}
Because of the relation between concurrence and entanglement of
formation it is, therefore, interesting to ask whether concurrence
can be generalized to larger quantum systems. Indeed attempts have
been made to generalize the definition of concurrence to higher
dimensional composite systems \cite{uhlmann,rungta1,auden,
fei1,fan,bad,li,bha}. Uhlmann generalized the concept of
concurrence by considering arbitrary conjugations acting on
arbitrary Hilbert spaces \cite{uhlmann}. His motivation is based
on the fact that the tilde operation on a pair of qubits is an
example of conjugation, that is, an antiunitary  operator whose
square is the identity. Rungta et al defined the so-called
I-concurrence in terms of a universal-inverter which is a
generalization to higher dimensions of two qubit spin flip
operation, therefore, the pure state concurrence in arbitrary
dimensions takes the form \cite{rungta1}
\begin{equation}\label{Icon}
C(\psi)=\sqrt{\langle\psi|{\cal S}_{N_{1}}\otimes {\cal
S}_{N_{2}}\left(\left|\psi\right>\left<\psi\right|\right)\left|\psi\right>}
=\sqrt{2(1-\rm{Tr}(\rho_A^2))}.
\end{equation}
Another generalization is proposed by Audenaert et al \cite{auden}
by defining a concurrence vector in terms of  a specific set of
antilinear operators. As pointed out by Wootters, it turns out
that the length of the concurrence vector is equal to the
definition given in Eq. (\ref{Icon}) \cite{woot2}. Albeverio and
Fei also generalized the notion of concurrence by using
invariants of local unitary transformations as \cite{fei1}
\begin{equation}\label{Feicon}
C(\psi)=\sqrt{\frac{N}{N-1}\left(I_0^2-I_1^2\right)}
=\sqrt{\frac{N}{N-1}\left(1-\rm{Tr}(\rho_A^2)\right)},
\end{equation}
which turns out to be the same as that of Rungta et al up to a
whole factor. In Eq. (\ref{Feicon}) $I_0$ and $I_1$ are two former
invariants of the group of local unitary transformations. As a
complete characterization of entanglement of a bipartite state in
arbitrary dimensions may require a quantity which, even for pure
states, does not reduce to single number
\cite{nielsen,vidal1,jon,vidal2,vidal3}, Fan et al defined the
concept of a concurrence hierarchy as $N-1$ invariants of a group
of local unitary for $N$-level systems \cite{fan}. Badziag et al
\cite{bad} also introduced multidimensional generalization of
concurrence. Recently, Li  et al used a fundamental representation
of $A_{N}$ Lie algebra and proposed concurrence vectors for a
bipartite system of arbitrary dimension as \cite{li}
\begin{equation}\label{ANcon}
{\bf C}=\left\{\left<\psi\right|(E_{\alpha}-E_{-\alpha})\otimes
(E_{\beta}-E_{-\beta})\left|\psi^\ast\right> \left|
\alpha,\beta\in \Delta^{+} \right. \right\},
\end{equation}
where $\Delta^{+}$ denotes the set of positive roots of $A_{N-1}$
Lie algebra. An extension of the notion of Wooters concurrence  to
multi-qubit systems is also proposed in Ref. \cite{bha}.

In this contribution, I generalize the notion of concurrence
vectors to arbitrary multipartite systems. The motivation is
based on the fact that Wootters concurrence of a pair of qubits
can be obtained by defining the tilde operation as
$|\tilde{\psi}\rangle=S\otimes S |\psi^\ast\rangle$ instead of
$|\tilde{\psi}\rangle=\sigma_y\otimes \sigma_y \left
|\psi^\ast\right>$, where here  $S$ is the only generator of
rotation group $SO(2)$ in such basis that
$(S)_{ij}=\epsilon_{ij}$ where $\epsilon_{12}=-\epsilon_{21}=1$
and $\epsilon_{11}=\epsilon_{22}=0$. Therefore, a natural
generalization of spin flip operation for arbitrary bipartite
systems  leads to a vector whose components are obtained by
employing tensor product of generators of the corresponding
rotation groups. A suitable generalization of the definition for
multipartite system is also proposed by defining concurrence
vector as a direct sum of concurrence subvectors in the sense that
each subvector corresponds to one pair of particles. Therefore,
it is proposed to use the norm of each subvector as a measure of
entanglement shared between the corresponding pair of particles. A
criterion for the separability of bipartite states is then given
as : a state is separable if and only if  the norm of its
concurrence vector vanish. For a multipartite systems the
vanishing of the concurrence vectors is necessary  but not
sufficient condition for separability.

The paper is organized as follows: In section 2, the definition of
concurrence vectors is given.  In section 3, the generalization of
the concurrence vector for multipartite system is proposed. Some
examples are also considered in this section. The paper is
concluded in section 4 with a brief conclusion.

\section{Concurrence vectors for bipartite pure states}
In this section, we give a generalization of the concurrence for
an arbitrary bipartite pure state. For motivation, let us first
consider a pure state $\left|\psi\right>\in {\mathbb C}^2\otimes
{\mathbb C}^3$ with the following generic form,
\begin{equation}
\left|\psi\right>=\sum_{i=1}^{2}\sum_{j=1}^{3}
a_{ij}\left|e_i\otimes e_j\right>,
\end{equation}
where $\left|e_i\right>\;(i=1,2)$ and $\left|e_j\right>\;
(j=1,2,3)$ are  orthonormal real basis of Hilbert space ${\mathbb
C}^2$ and ${\mathbb C}^3$ respectively. Of course by means of the
Schmidt decomposition one can consider $\left|\psi\right>$ as a
vector in a ${\mathbb C}^2\otimes {\mathbb C}^2$ Hilbert space,
but to see the main idea of the paper we do not use Schmidt
decomposition. It can be easily seen that the entanglement of
$\left|\psi\right>$ can be written as
$E_f(\psi)=H\left(\frac{1}{2}+\frac{1}{2}\sqrt{1-C^2}\right)$,
where concurrence $C$ is defined by
\begin{equation}\label{23con1}
C=2\sqrt{|a_{12}a_{23}-a_{13}a_ {22}|^2+|a_{11}a_{23}-a_{13}a_
{21}|^2+|a_{11}a_{22}-a_{12}a_ {21}|^2}.
\end{equation}
On the other hand, Eq. (\ref{23con1}) can also be written as
\begin{equation}\label{23con2}
C=\sqrt{\sum_{\alpha=1}^{3}|\langle\psi |{\tilde \psi}_\alpha
\rangle|^2},
\end{equation}
where $|{\tilde \psi}_\alpha \rangle$ are defined
by
\begin{equation}\label{23con3} |{\tilde \psi}_\alpha
\rangle=(S\otimes L_\alpha) |\psi^{\ast}\rangle,
\end{equation}
where $S$ is the only generator of two-dimensional rotation group
$SO(2)$ with matrix elements $(S)_{ij}=\epsilon_{ij}$ and
$L_\alpha$ with matrix elements $(L_\alpha)_{jk}=\epsilon_{\alpha
jk}$ denote three generators of an $SO(3)$ group. Here
$\epsilon_{ij}$ is defined by $\epsilon_{12}=-\epsilon_{21}=1$,
$\epsilon_{11}=\epsilon_{22}=0$ and $\epsilon_{\alpha jk}$ is
antisymmetric under interchange of any two indices and
$\epsilon_{123}=1$.

Similarly, for pure state $|\psi\rangle\in {\mathbb C}^2\otimes
{\mathbb C}^N$ with the generic form
\begin{equation}
\left|\psi\right>=\sum_{i=1}^{2}\sum_{j=1}^{N}
a_{ij}\left|e_i\otimes e_j\right>,
\end{equation}
entanglement $E(\psi)$ is obtained by Eq. (\ref{EoF2}) with the
following:
\begin{equation}\label{2Ncon1}
C=2\sqrt{\sum_{j<k}^{N}|a_{1j}a_{2k}-a_{2j}a_ {1k}|^2 }.
\end{equation}
It is straightforward to see that Eq. (\ref{2Ncon1}) can be
expressed as
\begin{equation}\label{2Ncon2}
C=\sqrt{\sum_{\alpha=1}^{N(N-1)/2}|\langle\psi |{\tilde
\psi}_{\alpha} \rangle|^2}.
\end{equation}
Here $|{\tilde \psi}_{\alpha} \rangle=(S\otimes
L_{\alpha})|\psi^\ast \rangle$, $\alpha=1,..., N(N-1)/2$, where
$L_{\alpha}$ are generators of $SO(N)$ group with matrix elements
$(L_{\alpha})_{kl}=(L_{[j_1j_2\cdots
j_{N-2}]})_{kl}=\epsilon_{[j_1j_2\cdots J_{N-2}]kl}$ where
$\alpha$ is used to denote the set of $N-2$ indices
$[j_1j_2\cdots j_{N-2}]$ with $1\le j_1 <j_2<\cdots <j_{N-2}\le
N$ in order to label $N(N-1)/2$ generators of $SO(N)$, and
$\epsilon_{j_1j_2\cdots j_{N}}$ is antisymmetric under
interchange of any two indices with $\epsilon_{12\cdots N}=1$. To
achieve  Eq. (\ref{2Ncon1}) from Eq. (\ref{2Ncon2}) we used the
following equations:
\begin{equation}\label{sumepsilon1}
\epsilon_{kl} \epsilon_{k^\prime l^\prime}= \delta_{k
k^{\prime}}\delta_{l l^{\prime}}- \delta_{k
l^{\prime}}\delta_{k^{\prime}l} ,
\end{equation}
\begin{equation}\label{sumepsilon2}
\sum_{1\le j_1 <j_2<\cdots <j_{N-2}\le N} \epsilon_{[j_1j_2\cdots
j_{N-2}]kl} \epsilon_{[j_1j_2\cdots j_{N-2}]k^\prime l^\prime}=
\delta_{k k^{\prime}}\delta_{l l^{\prime}}- \delta_{k
l^{\prime}}\delta_{k^{\prime}l}.
\end{equation}
Next, to generalize the above definition of concurrence for an
arbitrary bipartite pure state let $\left|\psi\right>$ be a pure
state in Hilbert space ${\mathbb C}^{N_1}\otimes {\mathbb
C}^{N_2}$ with following decomposition
\begin{equation}\label{N1N2psi}
\left|\psi\right>=\sum_{i=1}^{N_1}\sum_{j=1}^{N_2}
a_{ij}\left|e_i\otimes e_j\right>,
\end{equation}
Now we define concurrence vector ${\bf C}$ with components
$C_{\alpha\beta}$ as
\begin{equation}\label{N1N2Ccomp}
C_{\alpha\beta}=\langle\psi |{\tilde \psi}_{\alpha\beta} \rangle,
\qquad |{\tilde \psi}_{\alpha\beta} \rangle=(L_{\alpha}\otimes
L_{\beta})\left|\psi^\ast \right>.
\end{equation}
where $L_{\alpha}$, $\alpha=1,..., N_1(N_1-1)/2$ and $L_{\beta}$,
$\beta=1,..., N_2(N_2-1)/2$ are generators of $SO(N_1)$ and
$SO(N_2)$ respectively. Now the norm of the concurrence vector can
be defined as a measure of entanglement, i.e.,
\begin{equation}\label{N1N2con1}
C=|{\bf C}|= \sqrt{\sum_{\alpha=1}^{N_1(N_1-1)/2} \;\;
\sum_{\beta=1}^{N_2(N_2-1)/2}|C_{\alpha\beta}|^2}.
\end{equation}
By using Eq. (\ref{sumepsilon2}) we can evaluate concurrence in
terms of parameters $a_{ij}$ where we get
\begin{equation}\label{C2}
C=2\sqrt{\sum_{i<j}^{N_1}\;\;\sum_{k<l}^{N_2}|a_{ik}a_{jl}-a_{il}a_
{jk}|^2 }.
\end{equation}
It is clear that $C(\psi)$ is zero when $|\psi \rangle$ is
factorizable, i.e., $a_{ij}=b_{i}c_{j}$ for some $b_i,\; c_j\;
\in {\mathbb C}$. On the other hand, $C$ takes its maximum value
$\sqrt{2(N-1)/N}$ with $N=\min{(N_1,N_2)}$, when $|\psi \rangle$
is a maximally entangled state. It should be noted that the result
is the same as that obtained in  \cite{fei1}, up to a whole
factor, therefore it is also in accordance with the result
obtained from the definition given in  \cite{rungta1}. It should
also be mention that the result is equal to the bipartite pure
state concurrence that Badsiag et al have defined in terms of a
trace norm of the concurrence matrix \cite{bad}. Interestingly,
components of the concurrence vector given  in Eq.
(\ref{N1N2Ccomp}) are same as elements of the concurrence matrix
of Ref. \cite{bad}. As a matter of fact, the definition given in
Eq. (\ref{N1N2Ccomp}) for concurrence vectors is closely related
to the definition proposed in Ref. \cite{li}. Actually, all
bipartite generalization of the concurrence lead to Eq.
(\ref{C2}). However, our objective here is to generalize the
definition for multipartite systems.

\section{Concurrence vectors for multipartite systems}
In order to further generalize the concept of concurrence vector
to multipartite systems, let us first analyze the problem that
arises in the definition of the pairwise entanglement between the
particles. In Eq. (\ref{N1N2Ccomp}) $|\psi^{\ast}\rangle$ can
also be written as $|\psi^\ast\rangle=(K_{1}\otimes
K_{2})|\psi\rangle$ where $K_1$ and $K_2$ are the complex
conjugation operators acting in ${\mathbb C}^{N_1}$ and ${\mathbb
C}^{N_2}$ respectively. Although the action of the direct product
of two anti-unitary transformation $K_{1}\otimes K_{2}$ on a
general ket $|\psi\rangle \in {\mathbb C}^{N_1}\otimes {\mathbb
C}^{N_2}$ can be properly defined,  the combination of an
anti-unitary and a unitary transformation such as $K_{1}\otimes
K_{2}\otimes I_3$ cannot be properly defined on a general ket
$|\psi\rangle \in {\mathbb C}^{N_1}\otimes {\mathbb
C}^{N_2}\otimes {\mathbb C}^{N_3}$ except that $|\psi\rangle$ is
factorized as
$|\psi\rangle=|\psi_{12}\rangle\otimes|\psi_3\rangle $ where
$|\psi_{12}\rangle\in {\mathbb C}^{N_1}\otimes {\mathbb C}^{N_2}$
and $|\psi_{3}\rangle\in {\mathbb C}^{N_3}$. This ambiguity can be
removed in the Hilbert Schmidt basis \cite{horo1} of the
corresponding system with
\begin{equation}\label{rhoT12}
\rho^{T_{12}}=(K_1\otimes K_2\otimes I_3)\rho(K_1\otimes
K_2\otimes I_3),
\end{equation}
for any $\rho=|\psi\rangle\langle\psi|$, whether $|\psi\rangle$
is factorizable or not. In Eq. (\ref{rhoT12}) $\rho^{T_{12}}$ is
the partial transpose of $\rho$ with respect to particles $1$ and
$2$.

Now to generalize the concept of a concurrence vector to
multipartite systems, let us consider an m-partite pure state
$\left|\psi\right>\in {\mathbb C}^{N_1}\otimes {\mathbb
C}^{N_2}\otimes \cdots \otimes {\mathbb C}^{N_m}$ which in the
standard basis has the following decomposition:
\begin{equation}\label{mpartitepsi}
\left|\psi\right>=\sum_{i_1}^{N_1}\sum_{i_2}^{N_2}\cdots
\sum_{i_m}^{N_m} a_{i_1 i_2 \cdots i_m}\left|e_{i_1}\otimes
e_{i_2}\otimes \cdots \otimes e_{i_m}\right>.
\end{equation}
We first define the pairwise entanglement between the particles.
Let $\rho=|\psi\rangle\langle\psi|$ be the density matrix
corresponding to pure state (\ref{mpartitepsi}). With
$\rho^{T_{ij}}$, we denote the matrix obtained from $\rho$ by
partial transposition with respect to subsystems $i$ and $j$,
i.e.,
\begin{equation}
\rho^{T_{ij}}=\left(|\psi\rangle\langle\psi|\right)^{T_{ij}}.
\end{equation}
Next, we define $\Pi_{i=1}^{m}N_i(N_i-1)/2$ dimensional
concurrence vector ${\bf C}$ with components
$C^{\{ij\}}_{\alpha_i\alpha_j}$ as
\begin{equation}
C^{\{ij\}}_{\alpha_i\alpha_j}= \sqrt{ \langle
\psi|\tilde{\rho}^{\{ij\}}_{\alpha_i,\alpha_j}|\psi\rangle},
\end{equation}
where
\begin{equation}
\tilde{\rho}^{\{ij\}}_{\alpha_i,\alpha_j}=M^{\{ij\}}_{\alpha_i\alpha_j}\rho^{T_{ij}}
M^{\{ij\}}_{\alpha_i\alpha_j}
\end{equation}
with
\begin{equation}
M^{\{ij\}}_{\alpha_i\alpha_j}=I_1\otimes \cdots \otimes I_{i-1}
\otimes L_{\alpha_i} \otimes I_{i+1} \otimes \cdots \otimes
I_{j-1} \otimes L_{\alpha_j} \otimes I_{j+1} \otimes \cdots
\otimes I_m,
\end{equation}
for $1\le i < j \le m$, $\alpha_i=1,\cdots, N_i(N_i-1)/2$ and
$\alpha_j=1,\cdots, N_j(N_j-1)/2$. Here $I_k$ denotes the
identity matrix in the Hilbert space of particle $k$, and
$L_{\alpha_i}$ represents the set of $N_i(N_i-1)/2$ generators of
an $SO(N_i)$ group with the following matrix elements
\begin{equation}\begin{array}{c}
\left<k_i\right|L_{\alpha_i}\left|l_i\right>=(L_{\alpha_i})_{k_i
l_i}=(L_{[i_1 i_2 \cdots i_{N_i-2}]})_{k_i l_i}=\epsilon_{[i_1 i_2
\cdots i_{N_i-2}]k_i l_i},
\\
1\le i_1 < i_2< \cdots < i_{N_i-2} \le N_i,
\end{array}
\end{equation}
and $L_{\alpha_{j}}$ are generators of an $SO(N_j)$ group with
similar a definition. The concurrence vector ${\bf C}$ is defined
in such a way that it involves all two-level entanglement shared
between all pairs of particles. Moreover, we can consider vector
${\bf C}$ as a direct sum of elementary subvectors ${\bf
C}^{\{ij\}}$, i. e.
\begin{equation}\label{direcsum}
{\bf C}=\sum_{\bigoplus ij}{\bf C}^{\{ij\}},
\end{equation}
such that each subvector ${\bf C}^{\{ij\}}$ corresponds to pair
$i$ and $j$ of particles. Accordingly  the entanglement
contribution of pair $i$ and $j$ in the entanglement of
$\left|\psi\right>$ can be defined as the norm of the concurrence
subvector ${\bf C}^{\{ij\}}$, that is
\begin{equation}\label{Cij}
\begin{array}{rl}
C^{\{ij\}}= & |{\bf C}^{\{ij\}}|=
\sqrt{\sum_{\alpha_i=1}^{N_i(N_i-1)/2}\quad\sum_{\alpha_j=1}^{N_j(N_j-1)/2}
\langle
\psi|\tilde{\rho}^{\{ij\}}_{\alpha_i,\alpha_j}|\psi\rangle}
\vspace{3mm}  \\
= & \left\{\sum_{\{K\}}\sum_{\{L\}}\sum_{k_i <l_i}^{N_i} \sum_{k_j
<l_j}^{N_j}\right. \\
& \left.\left|a_{\{k_i, k_j, K\}} a_{\{l_i, l_j, L\}}-a_{\{k_i,
l_j, K\}}a_{\{l_i, k_j, L\}}-a_{\{l_i, k_j, K\}} a_{\{k_i, l_j,
L\}}+a_{\{l_i, l_j, K\}}a_{\{k_i, k_j,
L\}}\right|^2\right\}^{1/2},
\end{array}
\end{equation}
where in the last line we used the following equations:
\begin{equation}
\begin{array}{l}
\sum_{[\alpha_i]} \epsilon_{[\alpha_i]k_il_i}
\epsilon_{[\alpha_i]k_i^{\prime} l_i^{\prime}}= \delta_{k_i
k_i^{\prime}}\delta_{l_i l_i^{\prime}}- \delta_{k_i
l_i^{\prime}}\delta_{k_i^{\prime} l_i },
\\
\sum_{[\alpha_j]} \epsilon_{[\alpha_j]k_jl_j}
\epsilon_{[\alpha_j]k_j^{\prime} l_j^{\prime}}= \delta_{k_j
k_j^{\prime}}\delta_{l_j l_j^{\prime}}- \delta_{k_j
l_j^{\prime}}\delta_{k_j^{\prime} l_j}.
\end{array}
\end{equation}
In Eq. (\ref{Cij}), $\{k_i, k_j, K\}$ stands for $m$ indices such
that $k_i$ and $k_j$ correspond to sybsystems $i$ and $j$
respectively, and $K$ denotes the set of $m-2$ indices for other
subsystems. Also $\sum_{\{K\}}$ stands for summation over indices
of all subsystems except subsystems $i$ and $j$.

It can be shown that $C^{\{ij\}}$ is zero when $|\psi\rangle$ is
factorizable among index $i$ and the rest of the system, i.e.,
when there exist some $b_{k_i},\; c_{\{k_j,K\}}\in \mathbb{C}$
such that $a_{\{k_i, k_j,K\}}=b_{k_i}c_{\{k_j,K\}}$. A similar
statement is also correct when $|\psi\rangle$ is factorizable
among index $j$ and the rest of the system. Also when two
subsystems $i$ and $j$ are disentangled from the rest of the
system, i.e., $a_{\{k_i, k_j,K\}}=b_{k_i,k_j}c_{\{K\}}$ for some
$b_{k_i,k_j},\; c_{\{K\}}\in \mathbb{C}$ , Eq. ({\ref{Cij}) takes
the form of Eq. (\ref{C2}), as we expect. This feature of
$C^{\{ij\}}$ shows that it can be considered as the pairwise
entanglement between the subsystems $i$ and $j$.

Finally, the total concurrence of $\left|\psi\right>$ may be
defined as the norm of the concurrence vector ${\bf C}$, i. e.
\begin{equation}\label{Cmpartite}
\begin{array}{rl}
C= & |{\bf C}|=\sqrt{\sum_{1\le i < j \le m} \left|{\bf C}^{\{ij
\}}\right|^2}
\vspace{3mm}  \\
= & \left\{\sum_{1\le i < j \le m}
\sum_{\left\{K\right\}}\sum_{\left\{L\right\}}
\sum_{k_i<l_i}^{N_i} \sum_{k_j <l_j}^{N_j}\right. \\
& \left.\left|a_{\{k_i, k_j, K\}} a_{\{l_i, l_j,L\}}-a_{\{k_i,
l_j,K\}}a_{\{l_i, k_j, L\}} - a_{\{l_i, k_j, K\}} a_{\{k_i,
l_j,L\}}+a_{\{l_i, l_j,K\}}a_{\{l_i, l_j, L\}}
\right|^2\right\}^{1/2}
\end{array}
\end{equation}
It is clear that if $|\psi \rangle$ is completely separable, if
$a_{i_1 i_2 \cdots i_m}=a_{i_1}b_{i_2}\cdots$ for some
$a_{i_1},\;b_{i_2},\cdots \in \mathbb{C}$, then all
$C^{\{ij\}}_{\alpha_i\alpha_j}$ are zero and entanglement of the
state  becomes zero. This is, of course, the first condition that
any good measure of entanglement should satisfy. Furthermore, the
entanglement measure should be invariant under local unitary
transformation, and its expectation should not increase  under
local operation and classical communication (LOCC). First, it
should be mentioned that for bipartite systems, where Eq.
(\ref{C2}) gives  concurrence of the state, the above conditions
are satisfied. This follows from the fact that Eq. (\ref{C2}) is
(up to a whole factor) the concurrence which is introduced by
Albeverio et al in terms of local invariants \cite{fei1}. On the
other hand, it is also equal to two-level concurrence defined in
Ref. \cite{fan}. In \cite{fan}, Fan et al  have defined a
hierarchy of concurrence for bipartite pure states and have shown
that they are entanglement monotones, i.e.,  cannot increase
under LOCC. Now, what can we say about monotonicity of the
concurrence for multipartite states? For any state we define a
concurrence vector the norm of which quantifies entanglement of
the state. The definition is such that for multipartite case, the
concurrence vector can be regarded as a direct sum of concurrence
subvetors (Eq. (\ref{direcsum})) in such a way that the norm of
each subvetor denotes pairwise entanglement shared between
particles. However, it should be stressed that this pairwise
entanglement (Eq. \ref{Cij}) is not obtained through tracing out
the remaining subsystems, but its definition comes directly from a
generalization of the bipartite case. The definition is simple and
computationally straightforward, and allows the concurrence to be
read off from the state. Our conjecture is that such defined
measure for multipartite systems is non-increasing under LOCC;
however, its proof is still in progress.

Next, to demonstrate the nature of this measure, we give some
multi-qubit examples in the following.

First, let us consider a three-qubit system with GHZ-state:
$|GHZ_3\rangle=(|000\rangle+|111\rangle)/\sqrt{2}$. For this
state Eqs. (\ref{Cij}) and (\ref{Cmpartite}) give
$C^{\{ij\}}=1/\sqrt{2}$ (for all $i,j$), and $C=\sqrt{3/2}$
respectively.

As another example, let us consider the W-state and anti W-state
defined with
$|W_3\rangle=(|110\rangle+|101\rangle+|011\rangle)/\sqrt{3}$ and
$|{\widetilde
W}_3\rangle=(|001\rangle+|010\rangle+|100\rangle)/\sqrt{3}$,
respectively. For these states we obtain $C^{\{ij\}}=2/3$ (for all
$i,j$), and $C=2/\sqrt{3}$.

For
$|\psi\rangle=(|00\rangle+|11\rangle)/\sqrt{2}\otimes(\alpha_3|0\rangle+\beta_3|1\rangle)$
with $|\alpha_3|^2+|\beta_3|^2=1$ we obtain $C^{\{12\}}=1$,
$C^{\{13\}}=C^{\{23\}}=0$ and $C=1$. The above examples show that
$C(|GHZ_3\rangle)>C(|W_3\rangle)=C(|{\widetilde W}
_3\rangle)>C(|EPR_{12}\rangle\otimes|\psi_3\rangle)$. Actually,
in an m-qubit system, the generalized GHZ-state
$|GHZ_m\rangle=(|0^{\otimes m}\rangle+|1^{\otimes m}
\rangle)/\sqrt{2}$ has the maximum total concurrence equal to
$C(|GHZ_m\rangle)=\sqrt{m(m-1)}/2$.

Let us now consider two superposition states which are considered
in Ref. \cite{wei}. First consider a superposition of the W-state
and anti W-state, i. e. $|W{\widetilde
W}(s,\phi)\rangle\equiv\sqrt{s}|W\rangle+\sqrt{1-s}\;e^{i\phi}|{\widetilde
W}\rangle$. For this state, we obtain
$C^{\{ij\}}=\frac{2}{3}\sqrt{\frac{3}{2}s(s-1)+1}$ and
$C=\sqrt{3}C^{\{ij\}}$. This result is in complete agreement with
the entanglement that was obtained in \cite{wei} by using a
geometric measure. Consider now a superposition of the W-state
and the GHZ-state, i.e.,
$|GW(s,\phi)\rangle\equiv\sqrt{s}|GHZ\rangle+\sqrt{1-s}\;e^{i\phi}|{W}\rangle$.
In this case we obtain
$C^{\{ij\}}=\frac{1}{3\sqrt{2}}\sqrt{s(5s-4)+8}$ and
$C=\sqrt{3}C^{\{ij\}}$. For this state, the geometric measure of
Ref. \cite{wei} observes an entanglement which is dependent on
phase $\phi$, but our measure shows an entanglement independent
of phase $\phi$. However, in this example as far as $s$ is
concerned, the behavior of our entanglement measure is closely
related to the behavior of the geometric measure of Ref.
\cite{wei}.

\section{Conclusion}
In summary, we gave the definition of a concurrence vector and
proposed  to use its norm as a measure of entanglement. In the
case of a bipartite pure state, it is shown that the norm of the
concurrence vector leads to the other proposals of generalization
of concurrence. In the multipartite case, the concurrence vector
is regarded as the direct sum of concurrence subvectors, each one
is associated with a pair of particles, therefore, the norm of
each subvector is used as the entanglement contribution of the
corresponding pair. We argue that the definition is not
exhaustive in order to completely quantify entanglement, so the
result of the paper is a small step towards quantifying the
entanglement. Also the definition of concurrence vectors
considered in this paper is just for pure states, and the problem
of mixed states remains open.

\end{document}